\documentclass{PoS}
 
\newcommand{\ie}{\textit{i.e.}} 
\def\Journal#1#2#3#4{{#1}\,{#2}, #3 (#4);}  
\newcommand{\etal}{et al.}
\def\citep#1{\cite{#1}}
\newcommand{\tableline}{\hline}
\newcommand{\DT}{\ensuremath{\Delta{T}}}
\newcommand{\R}{\ensuremath{R}}
\newcommand{\p}{{p}}
\newcommand{\He}{{He}}
\newcommand{\C}{{C}}

\newcommand{\ApJ}{Astrophys. J.}
\newcommand{\AeA}{Astron. \& Astrophys.}
\newcommand{\PRL}{Phys. Rev. Lett.}
\newcommand{\PRD}{Phys. Rev. D}
\newcommand{\PRC}{Phys. Rev. C}
\newcommand{\ASR}{Adv. Space Res.}

\newcommand{\JGR}{J. Geophys. Res.}

\title{Time lag in cosmic-ray modulation and global properties of the Solar Cycle}
\ShortTitle{Global properties of the cosmic-ray modulation time lag}

\author{{Bruna Bertucci}\thanks{E-mail: {bruna.bertucci@pg.infn.it}}\\
Department of Physics and Earth's Science, Universit{\`a} di Perugia, and INFN - Perugia, I-06100 Perugia, Italy\\
}

\author{{Emanuele Fiandrini}\thanks{E-mail: {emanuele.fiandrini@pg.infn.it}}\\
Department of Physics and Earth's Science, Universit{\`a} di Perugia, and INFN - Perugia, I-06100 Perugia, Italy\\
}

\author{{Behrouz Khiali}\thanks{E-mail: {behrouz.khiali@ssdc.asi.it}}\\
INFN - Roma Tor Vergata, and Space Science Data Center - ASI, I-01100 Roma, Italy\\
}

\author{{Nicola Tomassetti}\thanks{E-mail: {nicola.tomassetti@pg.infn.it}}\\
Department of Physics and Earth's Science, Universit{\`a} di Perugia, and INFN - Perugia, I-06100 Perugia, Italy\\
}

\abstract{
  When entering the heliosphere, Galactic cosmic rays (GCRs) are influenced by magnetic turbulence and Solar wind disturbances, which cause the so-called "solar modulation" effect. Understanding the time-dependent relationship between the Sun's variability and GCR flux modulation is essential for the investigation of the GCR transport processes in the heliosphere, as well as for the establishment of predictive models of GCR radiation in the interplanetary space. The known anti-correlation between GCR flux and sunspot number appears to be delayed by several months, but the origin of such a time lag is unclear. In this work, we are perform the first global characterization of the time lag evolution over the solar cycles and its energy dependence. We made use of a large collection of time-resolved data, both from space missions and ground based observatories. Since the long-term variation of the GCR flux originates by a combination of several physics processes, the investigation presented here may reveal important aspects of the GCR transport in the heliospheric plasma.
} 
\FullConference{36th International Cosmic Ray Conference - ICRC2019 -\\
		July 24th - August 1st, 2019\\
		Madison, WI, U.S.A.}
\begin{document}

\section{Introduction}      
\label{Sec::Introduction}   

When entering the heliosphere, Galactic cosmic rays (GCRs) travel through a turbulent magnetized plasma
--the solar wind and its embedded magnetic field-- which reshapes their energy spectra.
As a result, the spectrum of GCRs observed near-Earth is significantly different from their 
local interstellar spectrum (LIS) beyond the boundaries of the heliosphere \cite{Potgieter2013}.
In particular, GCRs in the heliosphere are spatially diffused by the small-scale irregularities of the heliospheric magnetic field (HMF),
drifted along the large-scale regular HMF component, advected and adiabatically decelerated by the outward flowing solar wind.
These processes lead to time-dependent modifications of the GCR flux intensities and their energy spectra, which is known as \emph{solar modulation} effect. 
The temporal evolution of solar modulation follows a quasi-periodical behavior.
It appears to be well (anti)correlated to the 11-year Solar Cycle.
The cycle of solar activity is related with generation strong magnetic fields in the  interior of the
Sun and manifested by the periodical appearance of sunspots  on its surface. 
The observed sunspot number (SSN), also known as Wolf number and often reported on monthly basis,
is then a good proxy for the magnetic activity in the Sun, and it is in fact utilized to fully characterize
the phases of the 11-year Solar Cycle.
The SSN is also widely used for studying the correlation between solar activity and GCR modulation \cite{Usoskin1998,Ross2019}.
Understanding of solar modulation is of great importance, in the physics of GCRs, to infer their LIS's and to identify their sources,
or to study the physics of particle transport in magnetized plasmas.
Moreover, predicting the level GCR radiation near-Earth is essential for astronauts and the electronic
components radiation hazard in long-duration space missions.
In this respect, several models have been proposed in order to predict the GCR intensity and evolution, near-Earth or in the interplanetary space. 
The results of these studies can be applied in studies of long-term solar-terrestrial relations and the global evolution of the heliosphere.

A major challenge of these efforts is to establish reliable relations between the modulation effects on the GCR fluxes
and the changing conditions of the Sun's magnetic activity.
As we will discuss in this paper, the recent direct measurements of GCRs operated in space
reveal the presence of a eight-month lag between solar activity and near-Earth GCR flux variation.
Such a lag is an essential input for predictive models of GCR radiation in the interplanetary space
\cite{Kuznetsov2017,Matthia2013,Norbury2018,ONeill2015}. At this purpose, however, it is essential to
investigate the lag dependence upon the solar cycles. 
In this paper, we present our efforts to determine the GCR modulation time-lag over several solar cycles,
using long time-series data from Neutron Monitors (NMs) in order to investigate
the long-term dependence of this phenomenon over the evolving solar activity.

\section{Time lag from GCR measurements in space}  
\label{Sec::Model}                                 

To investigate the solar modulation effect of GCRs in the heliosphere, we made use of a large variety 
of GCR and solar data collected in space missions or in high-altitude balloons. 
In particular we made use of a large compilation of data on GCR protons \cite{Bindi2017}, 
We include data from EPHIN/SOHO, where GCR protons have been measured on yearly basis, between 1995 and 2016, at about 1 GeV of kinetic energy  \cite{Kuhl2016}.
Measurements of monthly fluxes of GCR protons have been reported recently by AMS, omboard the ISS since 2011,
at energy between 0.3 and 50 GeV/n \cite{Aguilar2018PHeVSTime}.
Previously, GCR proton (and helium) data from 2011 to 2013 were reported \cite{Aguilar2015Proton,Aguilar2015Helium}.
Monthly fluxes of GCR protons have also been measured by the PAMELA experiment between 2006 and 2014,
in the kinetic energy range 0.1-50 GeV/n \cite{Adriani2013,Martucci2018}.
We also include the data from the balloon based BESS Polar-I (Polar-II) mission,
collected from 13 to 21 December 2004 (23 December 2007 to 16 January 2008) \cite{Abe2012}. 
Solar data include the
Wolf number (or SSN) and the tilt angle $\alpha$ of the heliospheric current sheet (HCS).
To model the SSN, we used a smoothed $\hat{S}(t)$-function that interpolates smoothly the monthly-series of SSNs provided
by the SILSO/SIDC database of the \emph{Royal Observatory of Belgium} \cite{CletteLefevre2016}. 
To model the tilt angle, we adopted a smooth interpolation  $\hat{\alpha}(t)$ of the time-series
provided on a 10-day basis by the \emph{Wilcox Solar Observatory} (radial model) \cite{Hoeksema1995}.

As input for the calculations, we need the GCR proton spectrum in interstellar space, $J^{\rm IS}_{p}$.
We calculated the proton LIS using improved models of GCR acceleration and
transport \cite{Tomassetti2018PHeVSTime,Tomassetti2015TwoHalo,Tomassetti2015PHeAnomaly,Tomassetti2012Hardening,Feng2016}.
Using direct LIS measurements from Voyager-1, in combination with high-energy data from
AMS \cite{Cummings2016,Aguilar2015Proton,Aguilar2015Helium}, we derived tight constraints for our GCR proton LIS model.

The GCR transport  in the heliosphere is modeled using our stochastic numerical framework \cite{Tomassetti2017TimeLag}.
The model provides the near-Earth GCR spectrum $J(t,E)$ at some epoch $t$, once the steady-state LIS $J^{\rm IS}(E)$ is specified
In particular, our model accounts for several GCR physics processes such as  diffusion, convection, adiabatic cooling,
and drift motion, that are described by a set of time-dependent parameters. Details on the physics modeling are
given elsewhere \cite{Tomassetti2017TimeLag,Tomassetti2017BCUnc,Kappl2016,Strauss2017}.

As in \cite{Tomassetti2017TimeLag}, we introduce the \emph{time lag} parameter \DT{} in the equations,
defined as a time delay in our calculations for GCR transport inside the heliosphere.
In practice, the lag appears in the relation between model parameters and solar-physics inputs, \ie, SSN and tilt angle.
For instance, the normalization factor of the GCR diffusion tensor $k^{0}$, is related to the smoothed SSN
via the function $\kappa^{0}(t) = a+b\log(\hat{S}(t-\DT))$, where $a$, $b$, and $\DT$ are free parameters.
Similarly, a \emph{retarded} tilt angle $\hat{\alpha}(t_{j}-\DT)$ is used as model input, to compute the GCR flux at a given epoch $t$.
This approach of using retarded functions for the solar indices (SSN and $\alpha$)
reflect our interpretation of the lag: a finite amount of time needed by the wind
to transport the Sun's perturbations to the outer heliosphere.
If we regarding the heliosphere as a spherical bubble-like region, with radius $d$$\sim$100-120\,AU
and with a radially flowing wind of speed $V$$\sim$300-700\,km/s,
we then expect a time-lag of the order of \DT$\sim$0.5-1\,year.

In our previous work
this approach led to the evidence for a time lag of 8.1 months.
The results were based on 3993 proton data points, at kinetic energy between 0.08 and  50\,GeV,
collected between 2000 and 2012 (during $A<0$ conditions in solar cycles 22-23).
In the work presented here, we have included the new GCR proton data
reported by AMS \cite{Aguilar2018PHeVSTime} and PAMELA \cite{Martucci2018}.
However, in the global fit, we consider the same period between 2000 and 2012 ($A<0$ polarity).
%
\begin{figure*}[!t]
\centering
\includegraphics[width=0.92\textwidth]{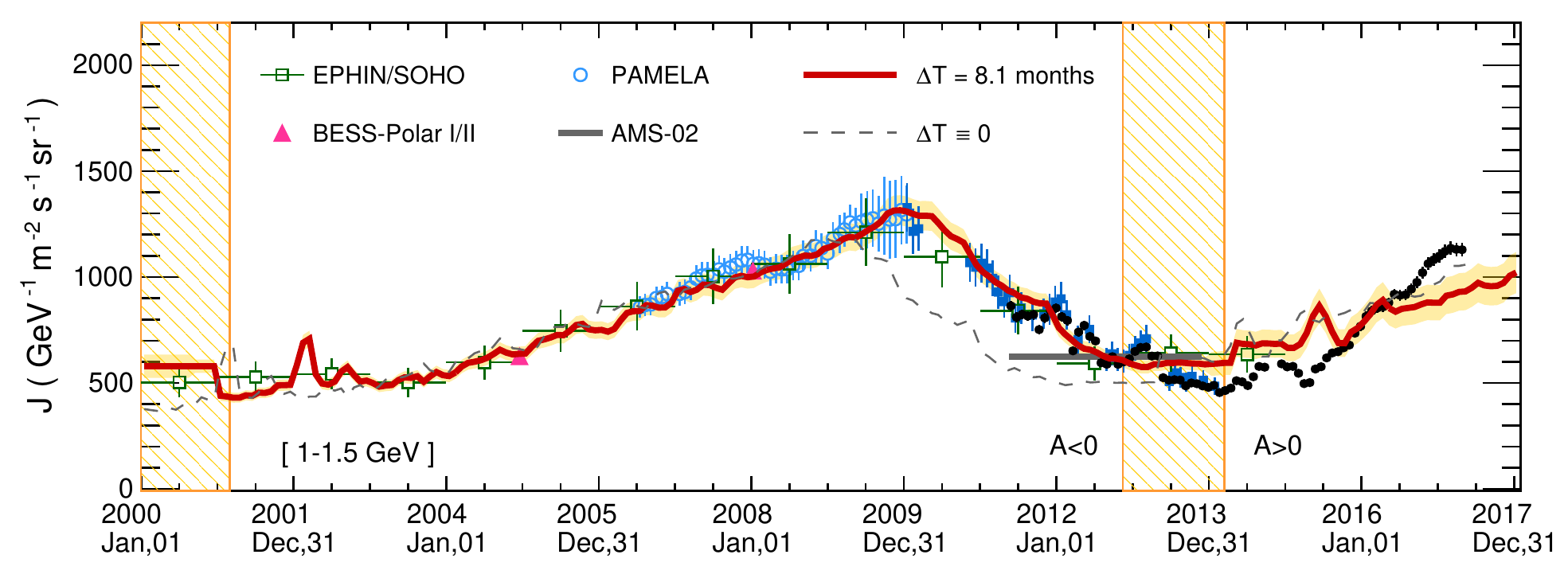}
\caption{\footnotesize{%
Time profile of the proton flux at $E=1-1.5$\,GeV. Best-fit calculations are shown as thick solid line,
along with the uncertainty band, in comparison with the data \cite{Adriani2013,Kuhl2016,Abe2012,Aguilar2015Proton}. 
In particular, the new data from PAMELA \cite{Martucci2018} (filled blue square) and from AMS \cite{Aguilar2018PHeVSTime} (black points)
between 2009 and 2017 are shown. Calculations for $\Delta{T}=0$ are shown as thin dashed lines. 
The shaded bars indicate the magnetic reversals of the Sun's polarity \cite{Sun2015}.
}}
\label{Fig::ccProtonTimeProfile}
\end{figure*}
%
%
The best-fit results are shown in Fig.\,\ref{Fig::ccProtonTimeProfile} for the temporal evolution of the GCR flux at energy $E{\approx}1.25$\,GeV.
In the figure, best-fit model calculations are shown comparison with the data under the \emph{delayed} model (thick red line)
and the \emph{unretarded} model (dashed black line). All fits are performed to data under $A<0$ polarity.
From the figure, it can be seen that the model with eight-month lag describes the data much better.
The plot of Fig.\,\ref{Fig::ccProtonTimeProfile}, however, shows that the predictions for the $A>0$ period
(after the 2013 polarity reversal), do not agree well with the AMS data. 
From the data, the post-reversal proton flux increases with time, like the predicted one,
but the increasing rate and the shape of the behavior are not well reproduced by the model.
Such a discrepancy suggests that some important physical inputs may be missing in the model. 
For instance, the use of a unique value for the GCR time lag could represent an oversimplification,
because the lag is determined using only GCR data from a specific phase of a solar cycle (in particular, with negative polarity).
From other NM based work, it was reported a odd-even effect for the time lag derived from different cycles.
More in general, one may expect a cycle-, phase-, or polarity-, dependence for the lag.
To investigate these suggestions, however, it is necessary to cover a large time period, and possibly including several solar cycles.
Thus, the collection of NM data is very useful for this investigation.

\section{Time lag studies from NM data}  
\label{Sec::TimeLagFromNM}               

To study the GCR modulation time-lag over a large period of time, and possibly including
many solar cycles, we use data from the worldwide network of ground-based NM detectors.
NM data consist in energy-integrated counting rates, $\mathcal{R}_{\rm NM}(t){\equiv}dN/dt$,
corrected for detector efficiency and atmospheric pressure.
More precisely, NMs are counting detectors that measure the arrival rates of secondary particles
of  hadronic  showers  generated  by  GCR interactions  with  the  atmosphere \cite{Dorman2009,Ghelfi2016,Usoskin2011}. 
In comparison with direct measurements of the GCR fluxes (that are energy-, particle-, and time-resolved) NMs give a poor information,
as they are only sensitive to the time evolution of the \emph{total} GCR flux, \ie, integrated over all relevant
energies and over all contributing GCR particles.

To model the NM response under a GCR transport model, we make use of a simple
force-field approximation (FFA). The FFA model is often applied to NMs,
as the typical GCR energies contributing to the NM signals are higher than 10\,GeV.
Within the FFA, the modulated GCR flux at epoch $t$ is related to its LIS by the modulation potential $\phi$: 
\begin{equation}\label{Eq::ForceField} 
J(t,E) = \frac{(E+ m_{p})^{2}- m_{p}^{2}}{\left( E+m_{p} +\frac{|Z|}{A}\phi \right)^{2}-m_{p}^{2}} \times J^{\rm IS}(E + \frac{|Z|}{A}\phi(t))
\end{equation}
where $Z$ and $A$ are the charge and mass numbers of the GCR particle, and $m_{p}$ is the proton mass.
Essentially, FFA consists into a kinetic energy shift $\Delta{E}=E^{\rm IS}-E$, for
GCR particles, to an average amount of $\Delta{E} - \frac{|Z|}{A} \phi$.
%
%
\begin{figure*}[!t]
\centering
\includegraphics[width=0.92\textwidth]{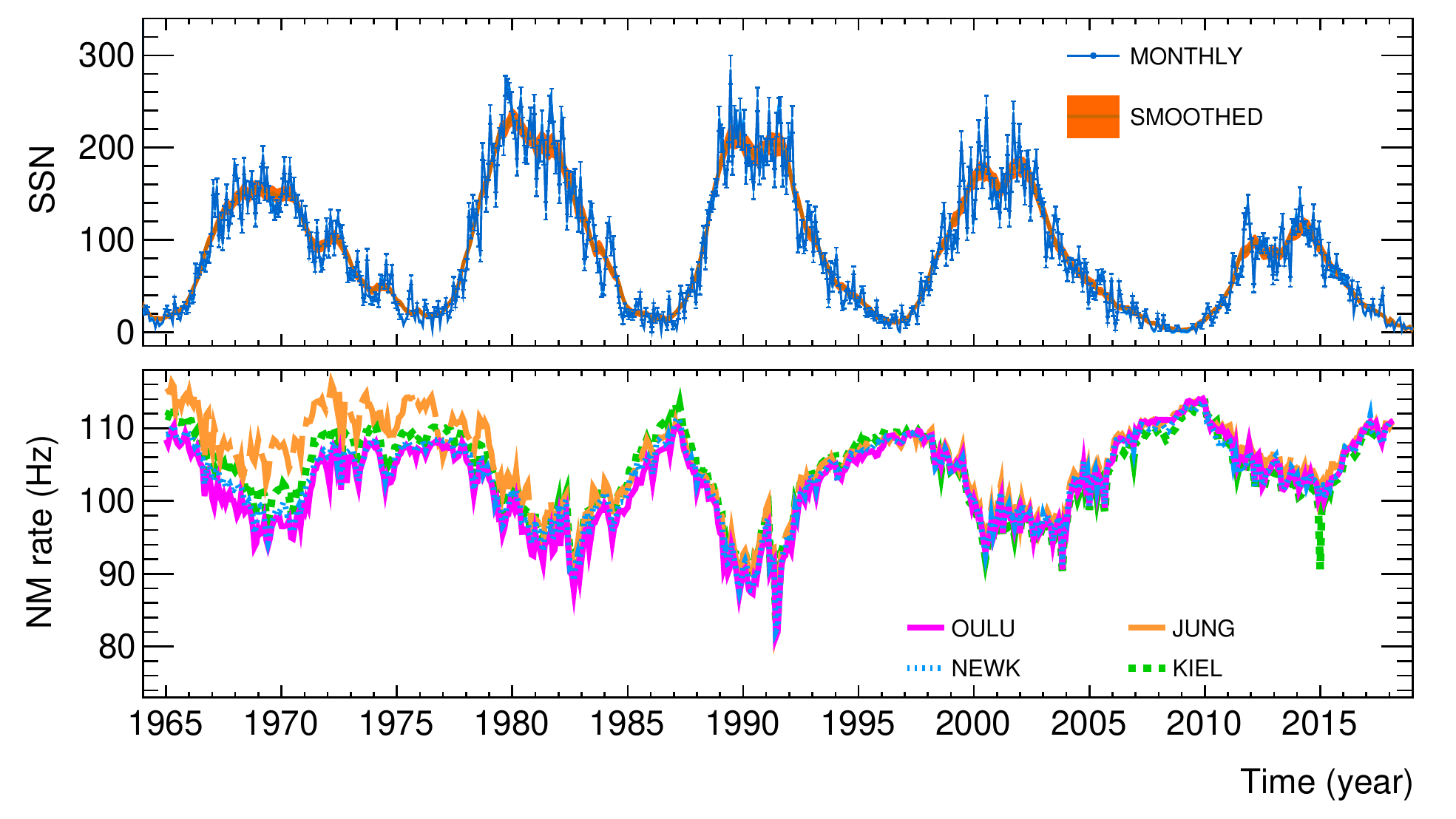}
\caption{\footnotesize{%
    Monthly SSN as function of time, from 1964 to present epoch, along with the NM counting
    rates from various stations. The anticorrelation between NM rates and SSN is apparent. 
}}
\label{Fig::ccSSNvsNM}
\end{figure*}
%
Within the FFA, we convert the monthly average NM rates $\mathcal{R}_{\rm NM}(t)$ into time-series of modulation potential $\phi=\phi(t)$.
This will allow us to compare data from different NM stations and extract information on the GCR flux variation.
To extract the information, we modeled the response of NM detectors as in \citep{Tomassetti2017BCUnc}. 
For a given NM detector $d$, located altitude $h_{d}$ and geomagnetic cutoff $\R_{C}^{d}$,
the link between the counting rate $\mathcal{R}_{\rm NM}^{d}$ and top-of-atmosphere GCR fluxes $J_{j}$ (with $j=$\p,\,\He) is expressed by:
\begin{equation}\label{Eq::NMRate}
  \mathcal{R}_{\rm NM}^{d}(t) = \int_{0}^{\infty}  dE \cdot \sum_{j={\rm GCRs}} \mathcal{H}^{d}_{j}(E)\cdot\mathcal{Y}^{d}_{j}(t,E)\cdot J_{j}(t,E)
\end{equation}
where $\mathcal{H}^{d}$ is a \emph{transmission function} around the cutoff value $\R^{d}_{C}$,
described as a smoothed step function \citep{SmartShea2005,Tomassetti2015XS},
and $\mathcal{Y}^{d}_{j}$ is the $j$-particle dependent detector response function \citep{Ghelfi2016}.
We use a factorized form, $\mathcal{Y}^{d}_{j} = \mathcal{V}^{d} \mathcal{F}^{d}_{j}$,
where $\mathcal{F}^{d}_{j}(t,E)$ describes the time and energy dependence of the NM response, including the development
of hadronic showers \citep{Cheminet2013}, and the factor $\mathcal{V}^{d} \propto exp(f_{d}h_{d})$ set normalization and altitude dependence. 
The last factor  $J_{j}(t,E)$ represents the \emph{modulated} energy spectra of all contributing GCR species. They are relate to the LIS
by Eq.\,\ref{Eq::ForceField}. Here we consider $j=\p,\He,\C$, 
which accounts for nearly $99\,\%$ of the GCR flux.
For a given NM station $d$ at an epoch $t$, the parameter $\phi$ is then obtained from
the request of agreement between the \emph{measured rate} $\hat{\mathcal{R}}^{d}$ 
and the \emph{calculated rate} $\mathcal{{R}}^{d}$, from Eq.\,\ref{Eq::NMRate},
together with the requirement that $\int_{\Delta T^{d}}{\mathcal{R}}^{d}(t) dt = \int_{\Delta T^{d}} \hat{\mathcal{R}}^{d}(t) dt$
where the integration is performed over the total observation periods $\Delta T^{d}$ \citep{Tomassetti2017BCUnc}.
\begin{table*}[!t]
\begin{center}
\small
\begin{tabular}{ccccc}
\tableline
\tableline
NM station & \href{http://www01.nmdb.eu/station/newk/}{NEWK} & \href{http://www01.nmdb.eu/station/oulu/}{OULU} &  \href{http://www01.nmdb.eu/station/kiel/}{KIEL} & \href{http://www01.nmdb.eu/station/jung}{JUNG}  \\
\tableline
Detector type     & 9-NM64              & 9-NM64               & 18-NM64              & 3-NM64\\
Location          & Newark, Delaware    & Oulu, Finland        & Kiel, Germany        & Jungfraujoch, Switzerland\\
Coordinates       & 39.68\,N 75.75\,W   & 65.05\,N, 25.47\,E   & 54.34\,N, 10.12\,E   & 46.55\,N, 7.98\,E\\
Altitude          & 50\,m               & 15\,m                & 54\,m                & 3570\,m\\
Cutoff            & 2400\,MV            & 810\,MV              & 2360\,MV             & 4500\,MV \\
\tableline
\end{tabular}
\caption{Main characteristics of the NM stations used in this work \cite{Steigies2015}. \label{Tab::NMStations}}
\end{center}
\end{table*}
In this work, we have retrieved monthly-averaged measurements from several stations of the worldwide
NM network \cite{Steigies2015,Mavromichalaki2011}.
Here we consider NM stations in Newark, Oulu, Kiel, and Jungfraujoch.
For these stations we consider data from 1964 and 2019, that cover five solar cycles.
The key properties of all considered NM stations are illustrated in Table\,\ref{Tab::NMStations}. 

\begin{figure*}[!t]
\centering
\includegraphics[width=0.92\textwidth]{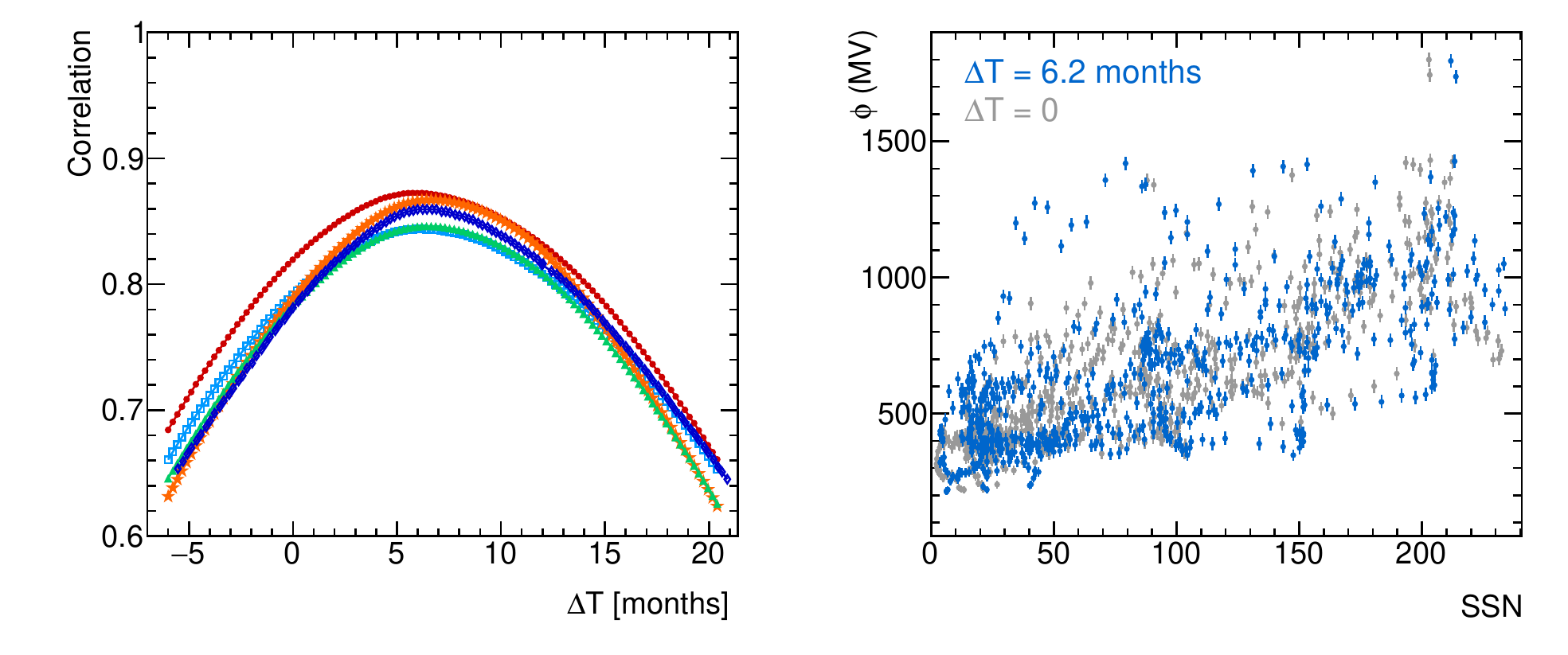}
\caption{\footnotesize{%
    Example of time-lag determination from the correlative analysis between monthly SSN and GCR modulation parameter,
    where the latter is obtained from NM rates (from the Kiel station, in the figure).
    The global lag from 1964 to 2019 is about six months.
}}
\label{Fig::ccSSNvsNM}
\end{figure*}

At this point, the time lag between GCR and solar data 
can be determined by means of a correlative analysis.
We analyzed the correlations  between the NM-driven modulation parameter $\phi$, at some reference epoch $t$,
and the smoothed SSN at the epoch $t-\DT$, under a defined time range.
For a given NM station $d$, the resulting lag \DT{} is determined as the parameter that maximizes 
the correlation between the modulation potential $\phi(t)$ (at epoch $t$) and the smoothed SSN $\hat{S}(t-\DT)$.
In practice, the correlation coefficient can be calculated as function of the lag \DT,
so that a fine scan over \DT will provide the \emph{best} estimation of the time lag.
A similar approaches were adopted in other works \cite{Zhu2018,Ross2019}.
As an example, in Fig.\,\ref{Fig::ccSSNvsNM} we show the determination of the overall time lag
estimated between 1964 and 2019, \ie, over five solar cycles.
In the figure, the correlation coefficient functions $\rho(\DT)$ are shown for various NM stations (left).
It can be seen that the $\phi$-SSN correlation improves if positive lag are accounted.
In particular, the $\rho(\DT)$ function peaks at about $6$ months of lag.
In the right panel, the scatter plot between $\phi$ and SSN is shown with and without time lag, for the KIEL station.  
Clearly, the results may show little dependence on the analyzed period. In particular,
dependence on the solar cycle have been noted \cite{Ross2019,AslamBadruddin2015,MishraMishra2018,Nymmik1995}
Using a generalized version of this method, we are now performing a global analysis based on the lag dependence on the solar cycles,
or in the different phases of the 11-year activity cycles, or its dependence upon the 22-year cycle of magnetic polarity.
The results will be presented at the conference and published in a forthcoming paper.

\section{Acknowledgements}  
\label{Sec::Results}        

We acknowledge the support of ASI under agreement \emph{ASI-UniPG 2019-2-HH.0}.
In this work, we made use of the \emph{NMDB.eu} real-time database for NM measurements,
the \emph{ASI-SSDC Cosmic-Ray Database} of the Space Science Data Center at ASI,
and the \emph{SILSO} SSN database of the \emph{Solar Influences Data Analysis Center}
at Royal Observatory of Belgium.
We thank Ilya Usoskin for providing us with updated time-series of the
$\phi$ parameter, that is a valuable reference for our work.


\end{document}